\documentclass[
10pt,twocolumn,notitlepage,oneside]{revtex4}
\usepackage{rotating}
\usepackage{amsfonts,longtable}
\usepackage{amsmath}
\renewcommand{\baselinestretch}{1}
\def\Ms{$\textrm{M}_{\odot}$}
\def\MspcII{$\textrm{M}_{\odot}/\textrm{pc}^{2}$}
\def\MspcIII{M$_{\odot}$/pc$^{3}$}
\def\cmIII{cm$^{-3}$}
\def\KcmIII{K/cm$^{3}$}
\def\kms{$\textrm{km/s}$}
\def\Mpc{$\textrm{Mpc}$}
\def\kpc{kpc}
\def\pc{pc}

\def\H2{H$_{2}$}
\def\HI{HI}
\def\roH2{$\rho_{\textrm{H}_2}$}
\def\roHI{$\rho_{\textrm{HI}}$}
\def\MH2{M$_{\textrm{H}_2}$}
\def\MHI{M$_{\textrm{HI}}$}
\def\M{M\,}
\def\ic{IC\,}

\begin{document}

\title{The Pressure of an Equilibrium Interstellar Medium in Galactic Disks}
\author{\firstname{A.~V.}~\surname{Kasparova}}
\email[]{anastasya.kasparova@gmail.com}
%\homepage[]{Your web page}
%\thanks{}
%\altaffiliation{}
\affiliation{Sternberg Astronomical Institut. }
\author{\firstname{A.~V.}~\surname{Zasov}}
\email[]{zasov@sai.msu.ru}
%\homepage[]{Your web page}
%\thanks{}
\affiliation{Sternberg Astronomical Institut.}
%\noaffiliation
%
\begin{abstract}
Based on an axisymmetric galactic disk model, we estimate the
equilibrium gas pressure $P/k$ in the disk plane as a function of
the galactocentric distance $R$ for several galaxies (MW, \M33,
\M51, \M81, \M100, \M101, \M106, and the SMC). For this purpose, we
solve a self-consistent system of equations by taking into account
the gas self-gravity and the presence of a dark pseudo-isothermal
halo. We assume that the turbulent velocity dispersions of the
atomic and molecular gases are fixed and that the velocity
dispersion of the old stellar disk corresponds to its marginal
stability (except for the Galaxy and the SMC). We also consider a
model with a constant disk thickness. Of the listed galaxies, the
SMC and \M51 have the highest pressure at a given relative radius
$R/R_{25}$, while \M81 has the lowest pressure. The pressure
dependence of the relative molecular gas fraction confirms the
existence of a positive correlation between these quantities, but it
is not so distinct as that obtained previously when the pressure was
estimated very roughly \cite{blitz04I}, \cite{blitz06II}. This
dependence breaks down for the inner regions of \M81 and \M106,
probably because the gas pressure has been underestimated in the
bulge region. We discuss the possible effects of factors other than
the pressure affecting the relative content of molecular gas in the
galaxies under consideration.

\textit{Astronomy Letters, 2008, Vol. 34, No. 3, pp. 152-162.}

\end{abstract}
\maketitle
\section{INTRODUCTION}

Almost all of the active processes associated with star formation
take place in a relatively narrow gaseous layer embedded in the
stellar disk. The gaseous layer is inhomogeneous in density and
temperature and is more homogeneous in pressure, since a more
tenuous medium is simultaneously hotter. For example, our Galaxy
shows that the change in gaseous-layer thickness along the disk
radius can be well reproduced by assuming a hydrostatic equilibrium
(see, e.g., ~\cite{narayan}). The {\HI} distribution along the $z$
coordinate indicates that the gas density at small $z$ decreases
approximately as $\exp(-z^2/h_z^2)$, where $h_z$ is the vertical
scale height of the gas distribution, i.e., according to a law
expected for an isothermal gaseous layer inside a homogeneous
stellar disk, although there is a density excess at large $z$
compared to this simple law \cite{dickey}, \cite{ferriere}.

Below, we will assume the galaxy to be axisymmetric and the gaseous
disk to be in an equilibrium state in which its thickness is
determined by turbulent gas velocities (although it is obvious that
the equilibrium condition can be violated in the local regions
associated with active processes in the disk).

The mean equilibrium gas pressure at a given galactocentric distance
R near the disk plane plays a very important (if not crucial) role
in the transition of the gas to molecular form \cite{elmegreen93},
\cite{blitz04I}, \cite{blitz06II}, \cite{blitz06} è \cite{Wong}.
Since the stars are formed in the molecular gas layer, the star
formation pattern and rate also depend on the conditions of the gas
transition from one phase to another, \HI~$\leftrightarrow$~{\H2}.
Having roughly estimated the turbulent equilibrium pressure of the
gaseous layer in several disk galaxies, Blitz and Rosolowsky
\cite{blitz04I}, \cite{blitz06II} concluded that the relative
molecular gas fraction increases almost linearly with pressure.
However, this important conclusion needs to be tested.

To calculate the pressure in galactic disks, the hydrostatic
equilibrium and Poisson equations are commonly used and a number of
simplifying assumptions are made. For an infinite disk with a
vertical gas scale height much smaller than that for the stellar
disk, when the contribution from the spheroidal components to the
"vertical" potential gradient is disregarded, the pressure of the
medium can be expressed by the formula (see, e.g., \cite{blitz04I})
\begin{equation}P = (2G)^{0.5}\Sigma_{gas} v_{gas}(\rho_{star}^{0.5} +
(\frac{\pi}{4}\rho_{gas})^{0.5}),
\end{equation}
where $G$ is the gravitational constant, $\Sigma_{gas}$ is the total
surface density of the gas, $v_{gas}$ is the velocity dispersion of
the gas in $z$ coordinate, and $\rho_{star}$ and $\rho_{gas}$ are,
respectively, the volume densities of the stellar and gaseous
components in the disk midplane. For a selfgravitating isothermal
stellar disk, the stellar surface density $\Sigma_{star} =
2\rho_{star}h_{star}$, where the vertical scale height of the
stellar disk is $h_{star} = (v_{star}^2/2\pi G \rho_{star})^{0.5}$.
Hence follows a simple formula for the equilibrium turbulent
pressure if the gas self-gravity is disregarded:
\begin{equation}
P =
0.84(G\Sigma_{star})^{0.5}\Sigma_{gas}\frac{v_{gas}}{h_{star}^{0.5}}.
\label{Press2}
\end{equation}
This expression is commonly used to calculate the pressure in the
plane of galactic disks \cite{blitz04I}, \cite{blitz06II},
\cite{Leroy}. Note that the stellar disk thickness cannot be
measured directly and it is generally taken rather arbitrarily. At
the same time, $v_{gas}$ and $h_{star}$ are assumed to depend weakly
on the galactocentric distance $R$ and to change little from galaxy
to galaxy. To a first approximation, the gas velocity dispersion is
about the same in different galaxies unless the regions of intense
star formation or circumnuclear regions are considered. In this
case, the turbulent pressure is a function of only the stellar and
gas surface densities, i.e., \mbox{$P(R) \sim
\Sigma^{0.5}_{star}(R)\Sigma_{gas}(R)$}. In this approach, not only
the change in stellar disk thickness with $R$, but also the
gravitational field of the gas and the dark halo is ignored. Blitz
and Rosolowsky \cite{blitz04I} argue that this approach gives a
pressure estimate with an accuracy of about 10\% for
\mbox{$\Sigma_{star}>20$~\MspcII}. However, as we will show below,
this error was underestimated significantly.

In this paper, we estimate the equilibrium turbulent pressure of the
interstellar medium that corresponds to the mean gas density in the
galactic plane at a given $R$ for several disk galaxies. We obtain
our estimates through a self-consistent solution of the equations
that describe the vertical volume density distributions of the
stellar, atomic, and molecular components of a disk that is assumed
to be axisymmetric. We take into account the gas self-gravity, the
possible change in stellar disk thickness with $R$, and the
contribution from the dark halo to the gravitational potential of
the galaxy.

\section{THE SAMPLE OF GALAXIES AND ADOPTED PARAMETERS}

\renewcommand{\baselinestretch}{1}
\begin{table*}
 \footnotesize
\begin{tabular}{lcccccccccccc}
\hline \hline \small
Name &Distance&$R_{25}$&   \multicolumn{3}{c}{Stellar disk}  &     {\HI}          & {\H2}    & \multicolumn{3}{c}{Dark halo}           & $V_{rot}$  &$v_z$       \\
\cline{4-6} \cline{9-11}
     &    &        &$h_r$&$\Sigma_{star}(0)$& Ref.          &     Ref.        &      Ref.       &  $V_{as}$   &   $R_c$   &   Ref.          &  Ref.      & Ref.       \\
     &{\Mpc} &  {\kpc}   &{\kpc}&{\MspcII}&&                 &                 &  {\kms}       &   {\kpc}     &                 &            &            \\
\hline
(1)  &(2) &(3)     &(4)  &(5)                           &(6)&(7)              &(8)              &(9)          &(10)       &(11)             &(12)        &(13)        \\
\hline
MW   & ---&  20    &  3.20 & 640.9          &\cite{narayan} &\cite{wolfire}   &\cite{wolfire}   & 220.0       &   5.0     &\cite{mera}      &     ---    &\cite{lewis}\\
\M33  & 0.7&  7     &  1.18 & 439.5          &\cite{regan}   &\cite{corbelli00}&\cite{heyer}     & 136.6       &   7.0     &\cite{corbelli03}&\cite{sofue}&    ---     \\
\M51  & 8.4&  27.4  &  4.38 & ---            &\cite{BIMA}    &\cite{boissier04}&\cite{boissier04}& 120.0       &   3.25    & ---             &\cite{sofue}&    ---     \\
\M81  &3.63&  11.55 &  2.8  & 1709.5         & ---           &\cite{Rots}      &\cite{boissier04}& 88.0        &   4.6     & ---             &\cite{sofue}&    ---    \\
\M100 &17.0&  18.3  &  4.00 & ---            &\cite{boissier}&\cite{boissier}  &\cite{nishiyama} & 272.0       &   4.7     & ---             &\cite{sofue}&    ---     \\
\M101 &7.48&  23.80 &  4.6  & 628.8          & ---           &\cite{Wong}      &\cite{Wong}      & 236.0       &   5.2     & ---             &\cite{sofue}&    ---    \\
\M106 &6.95&  10    &  6.3  & 933.7          & ---           &\cite{boissier04}&\cite{boissier04} & 157.0       &   8.0     & ---             &\cite{sofue}&    ---    \\
SMC  &0.06&  3.47  &  1.4  & 168.0          & ---           &\cite{stanimirovich}     & ---             & ---         &   ---     & ---             &\cite{stanimirovich}&\cite{SMC_2}\\
\hline \hline
\end{tabular}
\caption{Columns (2) and (3) give the assumed distance to the object
in {\Mpc} and its photometric radius $D_{25}/2$ in {\kpc}; (4) and
(5) list stellar disk parameters: the radial disk scale length in
{\kpc} and the surface density corresponding to $R=0$,
$\Sigma_{star}(0)$, in units of {\MspcII}; (9) and (10) list dark
halo parameters: the asymptotic velocity in {\kms} and the core
radius in {\kpc}; all of the remaining columns give references to
the papers containing the corresponding characteristics.}
\label{tab1}
\end{table*}

To estimate the equilibrium gas pressure at various R, we must
specify the radial surface density distributions of the main
galactic components and the velocity dispersions $v_i(R)$ for the
stellar, atomic, and molecular disks. We chose several galaxies for
which fairly complete data on the gas distribution, brightness, and
rotational velocity are available in the literature: our Galaxy,
\M33, \M51, \M81, \M100, \M101, \M106, and the Small Magellanic
Cloud (SMC). Basic characteristics of the galaxies are given in the
table~\ref{tab1}. All of the estimates used were reduced to the
given distances to the objects.

For \M81, \M101, \M106, and the SMC, the disk parameters were found
by modeling the rotation curves (best-fit model); for the remaining
galaxies, the surface density distributions of the stellar disks
either were taken from available publications or were recalculated
from the radial brightness profile. The mass-to-light ratio
calculated from stellar models based on disk color indices
\cite{bell} was used to pass from brightness to surface density.

The estimation of the vertical stellar disk scale height
(half-thickness) needed to determine the gas pressure presents a
particular problem. For every galaxy, we used two models. In the
first model, which we take as the preferred one, the stellar disk
half-thickness for the sample galaxies (except for our Galaxy and
the SMC) was estimated by assuming that the velocity dispersion of
the old stars constituting the bulk of the disk mass was minimal to
ensure its dynamical stability. Strictly speaking, the
half-thickness estimates obtained in this way are minimal. However,
analysis of the observational data shows that the velocity
dispersion for most spiral galaxies is actually close to its
critical value (see, e.g., \cite{bottema}, \cite{zasov02},
\cite{zasov}). In the second, simpler model, the stellar disk
thickness was assumed to be constant along $R$. The pressure
estimation procedure for this case is described in the next section.

To determine the half-thickness of a marginally stable stellar disk,
we initially found the epicyclic frequency from the model rotation
curve:
\begin{equation}
\ae(R) = 2\Omega\sqrt{1+\frac{R}{2\Omega}\frac{d\Omega}{dR}},
\end{equation}
where $\Omega(R) = V(R)/R$ is the angular velocity. For the
gravitational stability of a collisionless, infinitely thin
homogeneous disk with respect to axisymmetric perturbations, the
critical velocity dispersion according to the Toomre criterion is
\begin{equation}
C_{crit}(R) = \frac{3.36G\Sigma_{star}(R)}{\ae(R)}.
\end{equation}
In the general case, the critical radial velocity dispersion is
\begin{equation}
(v_r)_{star} = Q C_{crit}. \label{vr}
\end{equation}
The parameter $Q$ in our paper is assumed to be constant along the
radius and equal to 1.5. Simulations of marginally stable
collisionless galactic disks show that this value agrees well (to
within $\sim30\%$) with the numerical results in a large $R$
interval~--- except for the central region where the bulge dominates
and the outermost regions where $Q$ can be twice as high (see, e.g.,
\cite{khoperskov}). Note that an underestimation of $Q$ and, hence,
$(v_r)_{star}$ means an overestimation of the gas density and
pressure.

As residual velocity measurements for old disk stars show, the ratio
of the vertical and radial velocity dispersions for most spiral
galaxies lies within the range 0.5-0.8; for earlier-type galaxies,
this ratio is probably, on average, higher \cite{westpfall},
\cite{shapiro}. For all galaxies, we will use the approximate
relation:
\begin{equation}
(v_z)_{star} = 0.5(v_r)_{star}. \label{vz}
\end{equation}

For the old disk of our Galaxy and the SMC disk, the vertical
stellar velocity dispersions were taken from publications (see the
table~\ref{tab1} for references). For the SMC, observations point to
the velocity dispersion, $(v_z)_{star} = 27.5$~{\kms}, which changes
little with galactocentric distance and is close to the velocity
dispersion for the atomic gas ($22$~{\kms}). The latter value is
atypically high for the galaxies, which is probably due to the
interaction between the Magellanic Clouds. Since the abundance of
molecular hydrogen in this galaxy is low with respect to {\HI}, the
contribution from {\H2} to the pressure was ignored.

For all of the sample galaxies, except the SMC, we took constant
values of $(v_z)_{\textrm{{HI}}} = 9$~{\kms} and
$(v_z)_{\textrm{{H}}_2} = 6$~{\kms}. Other model parameters are the
core radius $R_c$ and the central volume density
$\rho_{\textrm{\tiny{DM}}}(0)$ (or asymptotic velocity $V_{as}$) for
a pseudo-isothermal galactic halo. For our Galaxy and \M33, these
parameters were taken from published data (see the table); for the
remaining spiral galaxies, these were found by modeling the rotation
curve. The profile of a pseudo-isothermal halo is described by the
formula
\begin{equation}
\rho_{\textrm{\tiny{DM}}}(r) =
\frac{\rho_{\textrm{\tiny{DM}}}(0)}{1+\left(r/R_c\right)^2}.
\end{equation}
In this case, the circular velocity is
$$V_c(R) = [4\pi
G\rho_{\textrm{\tiny{DM}}}(0)R_c^2(1-\dfrac{R_c}{R}\arctan(\dfrac{R}{R_c}))]^{0.5}.
$$
At large $R$, the velocity approaches its asymptotic value of
\begin{equation}
V_{as} = [4\pi G\rho_{\textrm{\tiny{DM}}}(0)R_c^2]^{0.5},
\end{equation}
therefore
\begin{equation}
\rho_{\textrm{\tiny{DM}}}(r) = \frac{V_{as}}{4\pi
G}\frac{1}{(R_c^2+r^2)}.
\end{equation}

Following Narayan and Jog \cite{narayan}, we took the core radius
for our Galaxy to be $R_c = 5$~{\kpc} and the asymptotic circular
velocity to be $V_{as} = 220$~{\kms} (as estimated by Mera et al.
\cite{mera}). The volume densities of the SMC components were
calculated without the bulge and dark halo, whose masses in this
galaxy are probably low \cite{stanimirovich}. In estimating the disk
parameters, we used the SMC rotation curve obtained from HI data and
corrected for the gas velocity dispersion (asymptotic drift), taken
from Stanimirovic et al. \cite{stanimirovich}.

\section{ESTIMATION OF THE EQUILIBRIUM PRESSURE}
\subsection{The System of Equations}

\begin{figure}
\includegraphics[width=7cm]{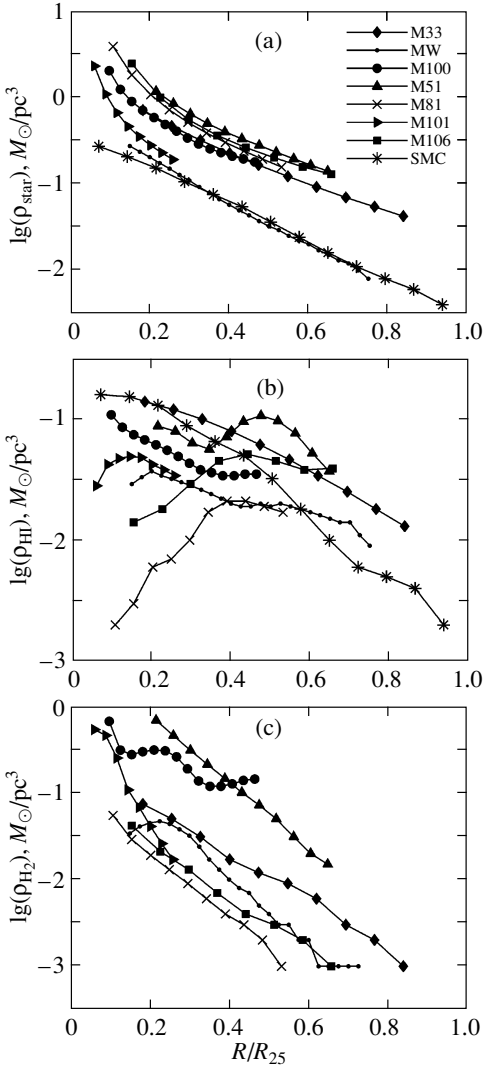}
\caption{Radial variations of the volume densities of the three
components in the galaxies of our sample. The densities are given in
units of {\MspcIII} and the Galactocentric distances are given in
fractions of the photometric radii $R_{25}$. \label{dens} }
\end{figure}

The thicknesses of the disk components, the midplane volume density
of the gas, and the corresponding pressure were estimated from the
atomic and molecular hydrogen surface densities, which were assumed
to be known. To calculate the thickness of a three-component disk
(stars, {\HI}, and {\H2}) in a general gravitational potential, we
used the same method as that applied by Narayan and Jog
\cite{narayan} for our Galaxy. We took into account both the
self-gravity of the individual components and the gravitational
influence of the halo, which can be significant in the outer disk
regions. The main simplification made in this case is that all disks
are assumed to be coplanar and axisymmetric, while the HI and H2
layers are assumed to be isothermal, i.e., the gas velocity
dispersions are constant with radius. Since the disk thickness is
much smaller than the radial density scale length, we ignored the
contribution from the radial density inhomogeneity to the potential
gradient in $z$. In a similar way, but using slightly different
input parameters and approximations, the disk thicknesses were
estimated by Abramova and Zasov \cite{abramova} for several galaxies
common to ours.

The basic hydrostatic equilibrium equation is
\begin{equation}
-\frac{\langle(v_z)^2_i\rangle}{\rho_i}\frac{d\rho_i}{dz}
=\sum_{i=1}^3\frac{\partial\phi_i}{\partial
z}+\frac{\partial\phi_d}{\partial z}
\end{equation}
where $\rho$ is the volume density, $-\partial\phi/\partial z$ is
the force per unit mass along the $z$ axis, $\phi$ is the
corresponding gravitational potential, the index $i$ pertains to one
of the three disk components (stellar disk, {\HI}, or {\H2}), and
the index $d$ pertains to the spherical halo.

The Poisson equation for a thin axisymmetric disk is
\begin{equation}
\sum_{i=1}^3\frac{\partial^2\phi_i}{\partial z^2} = 4\pi
G\sum_{i=1}^3\rho_i.
\end{equation}

The combination of the last two equations leads to an expression for
the volume density distribution at a given galactocentric distance:
\begin{equation}
\frac{d^2\rho_i}{dz^2} =
\frac{\rho_i}{\langle(v_z)^2_i\rangle}\left[-4\pi
G\sum_{i=1}^3\rho_i-\frac{\partial^2\phi_d}{\partial z^2}\right] +
\frac{1}{\rho_i}\left(\frac{d\rho_i}{dz}\right), \label{eq12}
\end{equation}
where the term in the first brackets corresponds to the potential of
the three-component disk inside the halo. The influence of the
galactic bulge on the disk thickness was disregarded, since the
volume density distribution, along with the velocity dispersions of
the disk gas and stars in the bulge region, are poorly known.
Therefore, the central regions of the galaxies were excluded from
our analysis.

The system of equations for the stellar and gaseous components was
solved numerically by a fourth-order Runge~--~Kutta method. The two
necessary boundary conditions in the $z = 0$ midplane of the disk
are
\begin{equation}
\rho_i = (\rho_0)_i\quad \textrm{and}\quad\frac{d\rho_i}{dz} = 0.
\end{equation}

The following normalization condition should be added to this: twice
the integral of the volume density over the $z$ coordinate for each
disk component should be equal to the surface (column) density
$\Sigma_i(R)$, which was assumed to be known in the R interval under
consideration.

The volume density of each of the three components was found by the
method of iterations. The first step is the simultaneous solution of
Eq.~(\ref{eq12}) for $\rho_{star}(z)$ with appropriate boundary
conditions at zero gas densities. Subsequently, the problem is
solved for $\rho_{\textrm{{HI}}}(z)$ using the stellar volume
density obtained at the previous step. At the next step, the system
is solved for $\rho_{\textrm{{H}}_2}(z)$ with the known
$\rho_{star}(z)$) and $\rho_{\textrm{{HI}}}(z)$. Indeed, this is not
enough, since the stellar disk was calculated without allowance for
the influence of the gas, while molecular hydrogen did not affect
the atomic gas density estimate. Therefore, we solved the system of
equations successively four more times using the densities
calculated in the previous iteration at each step. In this way, we
obtained self-consistent solutions for each component in the general
gravitational field and, hence, the volume density distributions
along $z$ at a given $R$.

\subsection{Pressure Variation Along the Galactic Radius}
\begin{figure*}[t]
\begin{center}
\includegraphics[width=15cm]{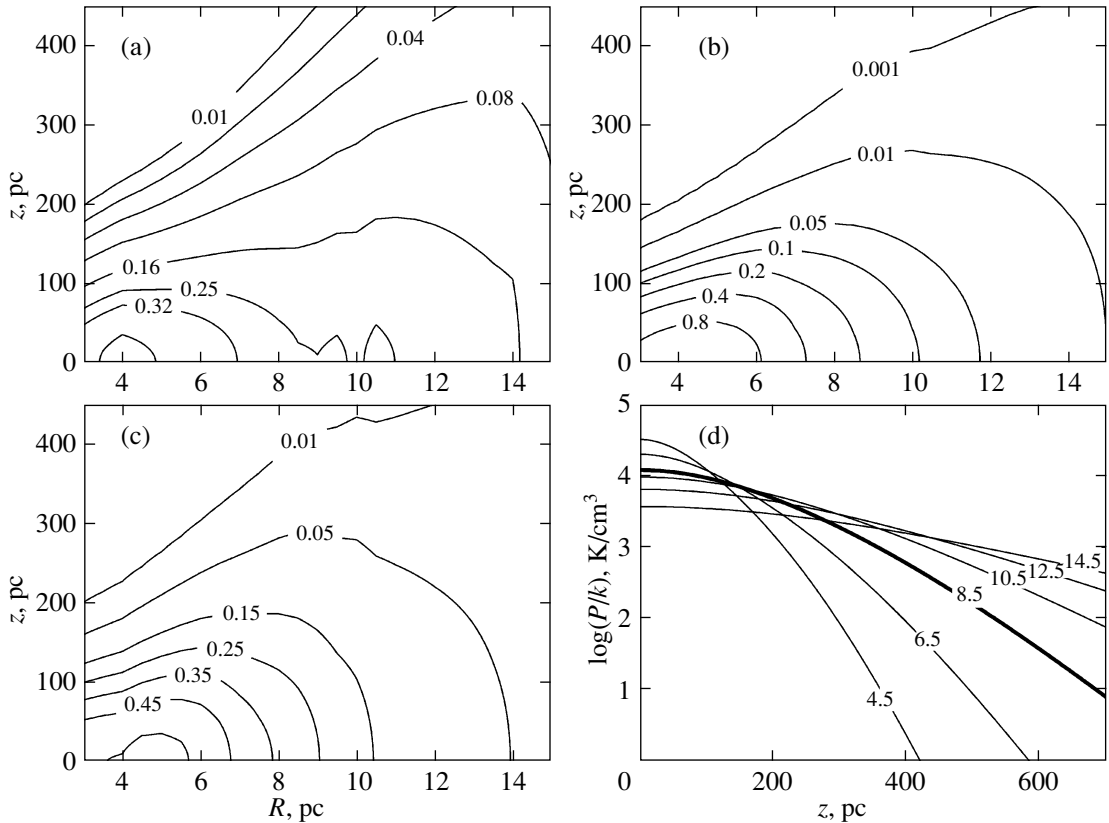}
\caption{Contour maps for the volume densities of the Galactic
gaseous components in units of {\cmIII}: (a) for the atomic gas, (b)
for the molecular gas, (c) the map of the molecular gas fraction
{\roH2/(\roHI+\roH2)}, and (d) pressure variations in z coordinate
at various Galactocentric distances; the heavy line represents the
curve for the solar distance.} \label{gas_contour}
\end{center}
\end{figure*}

The gas pressure was determined from the already obtained solutions
to the hydrostatic equilibrium and Poisson equations. To within the
errors of our iterative calculations (several percent), the
equilibrium pressure of the interstellar medium obtained during the
solution is equal to the dynamic pressure in the disk plane:
\begin{equation}
P = P_{dyn} =
\rho_{\textrm{HI}}v_{\textrm{HI}}^2+\rho_{\textrm{H}_2}v_{\textrm{H}_2}^2,
\end{equation}
where the contribution from elements heavier than hydrogen (the
coefficient 1.38) was taken into account in the gas densities.

We compared the radial pressure variations calculated by our method
and from the simplified dependence (\ref {Press2}) used by Blitz and
Rosolowsky \cite{blitz04I}, \cite{blitz06II}. Since these authors
took into account neither the radial variations in the thicknesses
of the disk components nor the gas self-gravity, the simplified
formula gives a systematic difference (underestimate) of the gas
pressure compared to our results. The discrepancy increases with
galactocentric distance and ranges from 30\% in the inner galactic
regions to more than 40\% at large $R$, because the gas and the dark
halo increase in importance in the outer disk regions.

The resulting estimates of the radial volume density variations for
the stellar and gaseous components in the midplane are illustrated
in Fig.\ref{dens}: (a) stellar densities, (b) atomic gas densities,
and (c) molecular gas densities. The SMC and MW have the lowest
stellar disk volume densities, while {\M81} and {\M106} have the
highest ones.We emphasize once again that the above density
estimates for all of the objects, except our Galaxy and the SMC,
were obtained by assuming a marginal gravitational stability of the
disk and, strictly speaking, give an upper limit for the density.

\subsection{Our Galaxy}

\begin{figure*}
\begin{center}
\includegraphics[width=13cm]{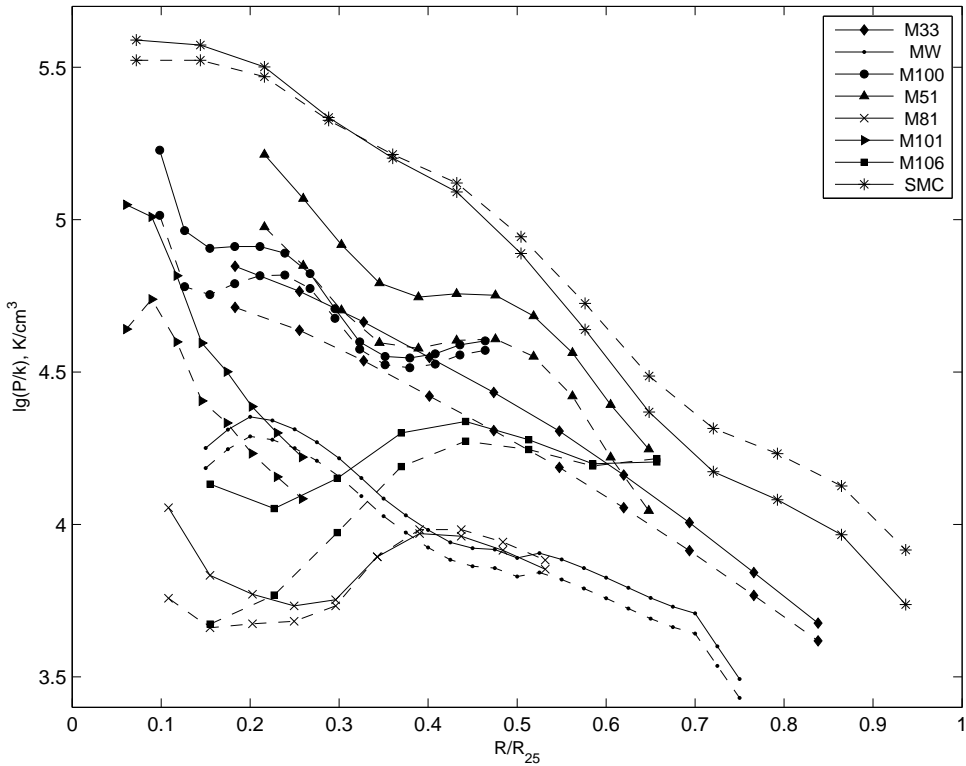}
\caption{Radial equilibrium pressure distribution for all our
galaxies. The solid and dashed lines represent, respectively, the
distributions obtained for our main model and by assuming the
stellar disk thickness to be constant.} \label{two_mod}
\end{center}
\end{figure*}

According to our calculations, the stellar volume density near the
solar orbit is $0.05$~{\MspcIII}, while the atomic and molecular gas
densities are, respectively, $0.02$ and $0.08$ in the same units,
corresponding to a total density of all disk components $\approx
0.15$~{\MspcIII}. The error in the stellar disk surface density,
which is known with an accuracy no higher than 30\%, introduces the
largest uncertainty into the estimate.

Our calculations allow the volume densities of the components to be
estimated at any point with coordinates $(R, z)$.
Figure~\ref{gas_contour} shows contours of the volume densities
{\roHI} (a) and {\roH2} (b). In agreement with the observational
data (see, e.g., \cite{imamura}), the atomic gas in our model forms
a layer that widens with galactocentric distance, although the
model, naturally, cannot reproduce the observed warp of the gas
layer at $R>12$ {\kpc}. The molecular layer is narrower than the
atomic one due to its lower velocity dispersion.

Figure~\ref{gas_contour}c presents a contour map that shows the
distribution of the molecular gas fraction with respect to the total
mass of the gas component, {\roH2/(\roHI+\roH2)}. We see from the
figure that this parameter decreases rapidly with z at
Galactocentric distances up to 7 {\kpc} and considerably more slowly
at larger distances. However, our model disregards the
molecular-to-atomic gas phase transition with increasing $z$.
Previously, Imamura and Sofue \cite{imamura} discussed the
possibility of this transition attributable to ultraviolet radiation
pressure and density variations. According to the model by these
autors, which is based on measuring the vertical molecular gas
density profile, the relative molecular fraction
{\roH2/(\roHI+\roH2)} decreases by half from its maximum value at a
distance of $\approx$ 80 {\pc} from the disk midplane, while
according to our model, this decrease takes place at $\approx$ 150
{\pc}. This confirms the possibility of a rapid, on a time scale of
less than $10^7$, transition of the gas from molecular to atomic
form and back. Numerical simulations show \cite{glover} that these
transitions can actually occur on a short time scale (several
million years) in the presence of supersonic turbulence.

\begin{figure*}[t]
\begin{center}
\includegraphics[width=15cm]{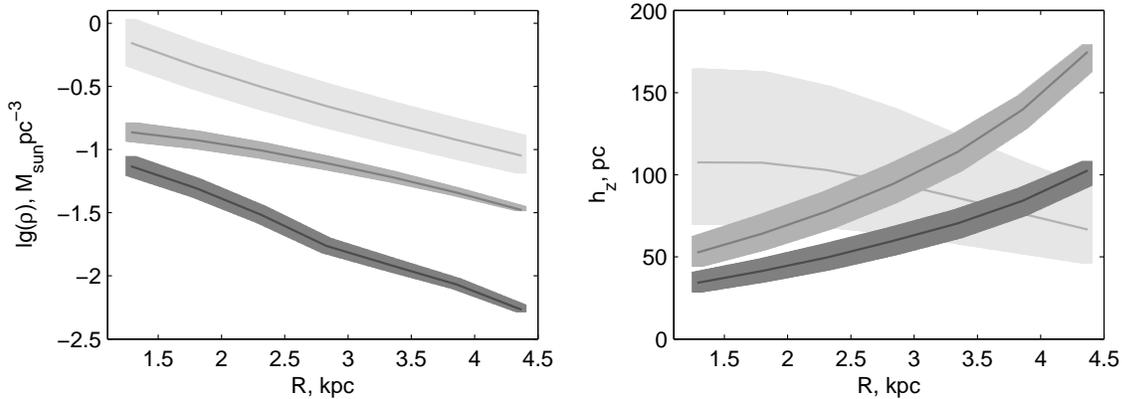}
\caption{Radial distributions of the volume densities for the three
disk components of {\M33} (left) and their vertical scale heights
(right). The dark-gray, light-gray, and lightest colors indicate
{\H2}, {\HI}, and stars, respectively. The variations in
$(v_z)_{star}$ by a factor of 1.25 relative to the value taken for a
marginally stable disk (solid lines) correspond to the shaded
regions.} \label{M33err}
\end{center}
\end{figure*}

Figure ~\ref{gas_contour}d illustrates the turbulent gas pressure
variations in $z$ coordinate at distances from 4.5 to 14.5 {\kpc} in
the disk plane at 2-{\kpc} steps. The heavy line indicates the
dependence for the solar distance (8.5 {\kpc}). This scheme clearly
illustrates a decrease in the vertical pressure gradient $|dP/dz|$
with Galactocentric distance. As a result, the gas pressure at small
$z$ is higher in the inner disk, but the reverse is true at large
$z$ and the pressure is higher in the outer Galactic regions.

\subsection{Radial Gas Pressure Profiles}

Parallel with the model of galaxies described above, where the
stellar disk thickness changes along $R$ and is determined by the
condition for its dynamical stability, we used a simpler model with
a constant disk thickness. Since the influence of the gravitational
fields from the gas and the dark halo on the stellar disk thickness
is ignored in this approach, the problem of estimating the densities
and half-thicknesses of the stellar and gaseous components ceases to
be completely self-consistent.

As the law $\rho_{star}(z)$, we took the formula for an equilibrium
gravitating isothermal disk

\begin{equation} \rho_{star}(z) = \rho_{star}(0)
\textrm{sech}^2\left(\frac{z}{(h_z)_{star}}\right),
\end{equation}
where the vertical disk scale height was assumed to be proportional
to its radial scale length

\begin{equation}(h_z)_{star} =\frac{(v^2_z)_{star}}{\pi G \Sigma_{star}} =
0.2h_R \label{hz}
\end{equation}
for all galaxies, except the SMC, and $0.3h_R$ for the SMC. The
half-thicknesses, densities, and pressures of the gaseous components
in the gravitational field of the stellar disk were determined at
the same gas velocity dispersions as those taken in the first (main)
model.

The equilibrium pressure distributions (the dependence of $P/k$ on
the radial coordinate normalized to the optical radius of the
galaxy, $R/R_{25}$) are shown in Fig.~\ref{two_mod} for the main
model (solid lines) and for the model with a disk of constant
thickness (dashed lines). When passing to the latter, the overall
pattern of the dependences for most galaxies, in general, changed
little; the difference between the pressure estimates does not
exceed a factor of one and a half ($\triangle lg(P/h) \geq $ 0.2),
except for the inner region of {\M106}, where $\triangle
lg(P/h)\approx 0.4$.

In general, the range of gas pressures in the galaxies under
consideration is almost two orders of magnitude. Of all the sample
galaxies, the SMC has the highest pressure, although the central
volume density of the stellar component for this object is lowest
(Fig.~\ref{dens}). The high pressure is related both to the high gas
content and to the high gas velocity dispersion in this galaxy. The
lowest pressure is in the outer parts of our Galaxy.

By varying the input model parameters, we investigated the degree of
their influence on the resulting distributions of the gas volume
densities and pressures as well as the vertical scale heights of the
stellar and gaseous components.

\begin{figure*}
\begin{center}
\includegraphics[width=15cm]{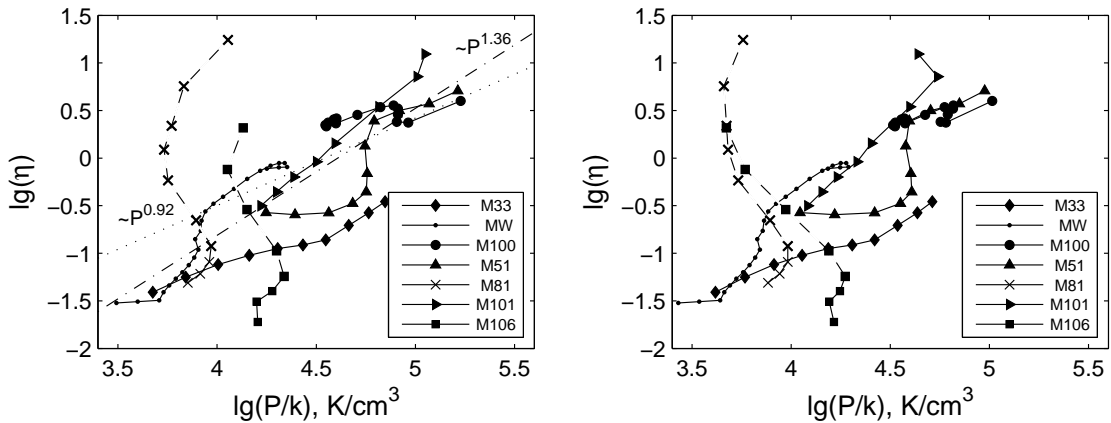}
\caption{Left: the dependence between of the molecular gas fraction
$\eta$ and the hydrostatic pressure calculated for a marginally
stable disk; the dash-dotted line reflects the average dependence
(without {\M81} and {\M106}), the dotted line indicates the straight
line corresponding to the correlation obtained by Blitz and
Rosolowsky \cite{blitz06II}. Right: the same dependence for the
model with a constant disk thickness, the dashed line marks the
inner regions of {\M81} and {\M106} (see the text).} \label{ETAtoP}
\end{center}
\end{figure*}

The solution turned out to be most sensitive to the adopted velocity
dispersion, primarily for the stellar disk $(v_z)_{star}$. For
illustration, Fig.~\ref{M33err} shows the radial distributions of
the volume densities for the three disk components and their
vertical scale heights (the dark-gray, light-gray, and lightest
colors correspond to {\H2}, {\HI}, and stars, respectively)
calculated, as described above, in terms of the model of a
marginally stable disk and after $(v_z)_{star}$ was increased and
decreased by a factor of 1.25 (shaded regions) for the galaxy
{\M33}. The central volume density of the stellar component
decreases approximately by a factor of 1.5 with increasing
$(v_z)_{star}$ and increases by the same factor with decreasing
$(v_z)_{star}$. However, the corresponding changes in gas density
and pressure are smaller, no more than 20\%. Obviously, the pressure
decreases with increasing $(v_z)_{star}$ and increases with
decreasing $(v_z)_{star}$.

\subsection{Relationship Between the Pressure and Molecular
Gas Content}

The pressure is assumed to play a dominant role in the
atomic-to-molecular gas phase transition (see the Introduction).
Based on the shielding condition for molecular clouds in an UV
radiation field, Elmegreen \cite{elmegreen93} found that the ratio
of the molecular gas volume density to the total gas density should
be proportional to $P^{2.2}j^{-1}$, where $j$ is the UV radiation
density. The simplified pressure estimates by Blitz and Rosolowsky
(\cite{blitz04I}, \cite{blitz06II}) discussed above led to the
conclusion that the relative molecular gas fraction $\eta =
\Sigma_{\textrm{H}_2}/\Sigma_{\textrm{HI}}\propto P^{0.8\div0.9}$
for galaxies with various star formation rates. Several galaxies
turned out to differ from others by a higher $\eta$ at the same
pressure. The authors suggested that this was related to the
interaction of the galaxies with the ambient medium. One of these
galaxies, {\M100}, belonging to the Virgo cluster, also enters into
our paper. However, according to our calculations, it exhibits no
anomalous behavior, which also put into question the mentioned
interpretation.

Figure~\ref{ETAtoP} shows the dependences $\eta(P/k)$ for the
galaxies of our sample using the two pressure calculation methods
described above: by assuming the stellar disk to be marginally
stable (left) and assuming its constant thickness (right).
Qualitatively, the two models yield similar results. All of the
galaxies, except {\M81} and {\M106}, definitely fall on the general
dependence: the amount of molecular gas compared to that of atomic
one also increases along with equilibrium pressure of the medium.
The dependence is best fitted by $\eta \sim P^k$ (dash-dotted line),
where $k = 1.34 \pm 0.44$ for the first model (without {\M81} and
{\M106}). For comparison, the dots in the figure mark the line that
describes the dependence taken from \cite{blitz04I}, $\eta \sim
P^{0.92}$. The disagreement with our relationship can result not
only from the rougher pressure estimates in the cited paper, but
also from the inclusion of galactic regions with very high
$\eta>10$, which are virtually absent in our dependences. As was
shown by Blitz and Rosolowsky \cite{blitz06II}, a linear dependence
more likely reflects the pressure estimation method than the
physical relationship between the quantities being compared for
$\eta>2$. The reason is that the pressure estimated from the
simplified formula is assumed from the outset to be proportional to
the atomic gas density and, at large $\eta>2$, to the molecular gas
density, which varies over a much wider range than the {\HI}
density.

The SMC is absent in the diagram: there is very little molecular gas
in this galaxy and it is distributed highly nonuniformly. Because of
the low heavy element abundance in the interstellar medium, the
conversion factor that relates the CO line intensity to the number
of {\H2} molecules on the line of sight for the SMC is probably much
higher than that commonly assumed for spirals, which makes it
difficult to determine the molecular gas mass. It is variously
estimated to be from several million {\Ms} to $3\cdot10^7$~{\Ms}
\cite{Leroy}, but it does not reach 0.1{\MHI} even in the latter
case, which is an order of magnitude smaller than the expected one,
given the high gas pressure (Fig.~\ref{two_mod}). This discrepancy
was also pointed out by Leroy et al. \cite{Leroy}. As in the case of
{\M33}, which also lies below the average line in the $\eta(P/k)$
diagram, it would be natural to associate this peculiarity with the
intensity of the UV radiation, which should play the most important
role in these two galaxies due to the low dust content and active
star formation.

Note the unusual behavior of {\M81} and {\M106} in the $\eta(P/k)$
diagram. It stems from the fact that in the inner regions of these
galaxies at distances of several {\kpc} from their cores, the gas
surface density ceases to increase or even decreases toward the
center (the corresponding segments of the curves for these galaxies
in Fig.~\ref{ETAtoP} are marked by dashes). For this reason, the
resulting pressure related to both gaseous components remains low.
However, since the {\H2} density does not exhibit the same deep
central ``dip'' as the {\HI} density, the molecular gas fraction
turns out to be high in this case. A similar peculiarity probably
also takes place in the Andromeda galaxy ({\M31}), which is not
among the galaxies considered here. Observations show that the {\HI}
density in this galaxy also decreases toward the center in the inner
disk, while the relative molecular gas fraction increases; this
effect probably cannot be explained by a change in conversion factor
\cite{M31}. Note also that although the conversion factor can be
lower in circumnuclear galactic regions due to the higher
metallicity in the gas, the three mentioned galaxies do not stand
out among the remaining ones by their relative oxygen abundance O/H
\cite{pilyugin}. The fact that the molecular gas content in the
inner regions of {\M33}, where the gas exhibits an underabundance of
heavy elements, nevertheless, follows the general dependence $\eta
(P)$ is indicative of the absence of a close correlation between the
gas metallicity and the degree of its molecularization. Similar
reasoning is given in \cite{blitz06II} for the galaxy {\ic10} as
well.

Another factor that can affect the {\H2} abundance and can lead to a
high $\eta$ in the inner regions of {\M31}, {\M81}, and {\M106} is a
low UV radiation density. At Galactocentric distances of several
{\kpc}, where $\eta$ is large, the intensity of the short-wavelength
radiation in {\M31} and {\M81} is actually low (see the GALEX images
of the galaxies), but this explanation is probably invalid for
{\M106}, where the inner region experiences a starburst.

An underestimation of the gas pressure by a factor of 3-10 due to
the action of some factors ignored in our simple model can in
principle be responsible for the unexpectedly high relative
molecular gas content in the inner regions of certain galaxies.
Since such galaxies have a large bulge (just as many of the galaxies
with a ring-like {\HI} distribution), it would be natural to
associate the high relative {\H2} mass precisely with its presence.

The bulge action on the gas is twofold. First, the bulge produces an
additional force that compresses the gas in the disk plane (in our
calculations, we took into account only the halo). To a first
approximation, the additional bulge-produced pressure is
\begin{equation}
P_b \approx \rho_{gas} g_z h_{gas}\approx \frac{1}{2}\sigma_{gas}
V_b^2 h_{gas}/R^2,
\end{equation}
where $g_z = GM_b h_{gas}$/$R^3$ is the $z$ acceleration component
attributable to the gravitational field of a bulge with mass $M_b$
and $V_b$ is the circular velocity component associated with the
bulge. It is easy to verify that at $h_{gas} \le 100$ {\pc} and $V_b
\le 200$ {\kms}, the pressure $P_b/k$ will not exceed significantly
$10^4$~\KcmIII. Therefore, including the bulge alleviates the
problem only slightly, but does not solve it.

The second possibility is the presence of a thermal hot plasma
associated with the bulge in which a high gas temperature is
supported by old stars (SNI explosions). The presence of a hot has
in the bulges of spiral galaxies manifests itself mainly as soft
X-ray emission. Although the observational data are so far rather
scarce, a gas with a temperature $(1\div7)\times 10^6$~K was
detected in the bulges and halos of several galaxies, including our
Galaxy and \M31 (see, e.g., \cite{wang}, \cite{yao}). Since the
pressure inside the cool gaseous layer cannot be lower than the
external pressure, a hot gas will play a crucial role at a gas
number density of $10^{-2}\div10^{-3}$~{\cmIII} in the bulge; this
can explain the pressure underestimation in the models that
disregard the influence of the ambient medium. Probably for the same
reason, the molecular gas fraction in lenticular galaxies within
several {\kpc} of the center is nevertheless high, despite the low
gas content in the disk (see, e.g., \cite{welch}).

Another factor that is disregarded in axisymmetric disk models is
the presence of spiral arms. The gas pressure inside of them is
higher and the molecular gas concentration to the spiral arms is a
well-established fact. The gas-compressing shock waves cause an
increase in the fraction of molecular gas and give rise to giant
molecular clouds. A tenuous molecular gas also exists between spiral
arms, although it is more difficult to detect (for a discussion, see
\cite{dobbs}). It is not yet clear how strong the influence of the
spiral pattern is on the degree of gas molecularization in the
galaxy as a whole. However, in any case, it by no means always plays
a significant role. Indeed, in contrast to the morphology of the
spiral pattern, the integrated ratios {\MH2/\MHI} in spiral galaxies
are almost independent of the morphological type or luminosity of
the galaxies \cite{boselli}. Spiral galaxies in which the bulk of
the gas is in molecular form are encountered among the galaxies of
all morphological types, except the latest ones \cite{kasparova}.
Note that the gas-rich \M51 and \M100 with both an extended spiral
pattern and a high relative molecular gas fraction $\eta$ stand out
among the galaxies considered in this paper by high fraction of
molecular gas. However, the latter agrees well with the higher
equilibrium gas pressure in these galaxies (see Fig.~\ref{ETAtoP}).

Thus, we have confirmed the dependence of the molecular gas fraction
on its mean equilibrium pressure at a given galactocentric distance,
although it is not so distinct as the dependence obtained when $P$
is estimated from very simplified formulas. However, such factors
disregarded in the model as the external pressure on the disk and
the intensity of the UV radiation can also play a significant role
in some galaxies.

\section{ACKNOWLEDGMENTS} This study was supported by the Russian Foundation
for Basic Research (project 07-02-00792).

% ñïèñîê ëèòåðàòóðû
%\newpage
\renewcommand{\baselinestretch}{1}
%\twocolumn

{\textit{Translated by V. Astakhov}
\end{document}